\documentstyle[prl,twocolumn,aps]{revtex}

\begin{document}
\def\beq{\begin{equation}}
\def\eeq{\end{equation}}
\def\bea{\begin{eqnarray}}
\def\eea{\end{eqnarray}}
\def\oupb{UPB\ }
\def\pb{PB\ }
\def\eps{\epsilon}
\newcommand{\ket}[1]{| #1 \rangle}
\newcommand{\bra}[1]{\langle #1 |}
\newcommand{\braket}[2]{\langle #1 | #2 \rangle}
\newcommand{\proj}[1]{| #1\rangle\!\langle #1 |}
\newcommand{\ba}{\begin{array}}
\newcommand{\ea}{\end{array}}
\newtheorem{theo}{Theorem}
\newtheorem{defi}{Definition}
\newtheorem{lem}{Lemma}
\newtheorem{exam}{Example}
\newtheorem{prop}{Property}


\draft

\title{Operational criterion and constructive checks for the separability \\of low rank density matrices.}

\author{Pawe\l{} Horodecki$^{1,2,}$\cite{poczta1},
Maciej Lewenstein$^{1,}$\cite{poczta2}, Guifr\'e Vidal$^{3}$\cite{poczta3}, 
and Ignacio Cirac$^{3}$\cite{poczta4} 
}
\address{$^1$ Institut f\"ur Theoretische Physik,
Universit\"at Hannover, D-30167 Hannover, Germany\\
$^2$ Faculty of Applied Physics and Mathematics,
Technical University of Gda\'nsk, 80--952 Gda\'nsk, Poland \\
$^3$ Institut f\"ur Theoretische Physik, 
Universit\"at of Innsbruck, A-6020 Innsbruck, Austria
}
\maketitle
\date{\today}

\begin{abstract}
We consider low rank density operators $\varrho$ supported on a $M\times N$ Hilbert space  for arbitrary $M$ and $N$ ($M\leq N$) and with a positive partial transpose (PPT) $\varrho^{T_A}\ge 0$. For rank $r(\varrho) \leq N$ we prove that having a PPT is necessary and sufficient for $\varrho$ to be separable; in this case we also provide its minimal decomposition in terms of pure product states. It follows from this result that there is no rank 3 bound entangled states having a PPT. 
We also present a necessary and sufficient condition for the separability of generic density matrices for which the sum of the ranks of $\varrho$ and $\varrho^{T_A}$ satisfies $r(\varrho)+r(\varrho^{T_A}) \le 2MN-M-N+2$. This separability condition has the form of a constructive check, providing thus also a pure product state decomposition for separable states, and it works in those cases where a system of couple polynomial equations has a finite number of solutions, as expected in most cases.
\end{abstract}

\date{\today}


\section{Introduction}
Entanglement is one of the quantum properties with no classical 
counterpart. It is closely connected to fundamental questions
of quantum mechanics\cite{EPR,Sch}, and to physical phenomena which are 
important for quantum information processing \cite{effects}.
The relevance of entanglement effects was first demonstrated for pure states.
However, in realistic physical situations one deals usually with
mixed states, in which pure state entanglement has been significantly weakened by noise. In order to overcome the problems caused by noise (i.e. in order to reduce it) the idea of distillation
of entanglement in spatially separated
laboratories was introduced \cite{Dist}. It was proved then\cite{pur}
that for bipartite systems of 
low dimensional Hilbert space ${\cal C}^M\times{\cal C}^N$ or simply $M\times N$ (namely systems with $M=2$ and $N=2$ or $3$) mixed state entanglement can always be distilled into its pure form. However, it turned out that in higher dimensional systems ($M N>6$) bound entanglement \cite{bound} -- which cannot be distilled, as opposed to free entanglement -- exists.

Unlike in the case of pure states, it is in general very difficult
to know whether a given mixed state is entangled 
(inseparable) or non-entangled (separable).
According to the definition a state supported on a Hilbert space ${\cal H}_{AB}={\cal H}_{A} \otimes 
{\cal H}_{B}$  
is separable if and only if it can be 
written in 
(or approximated by) the form \cite{Werner}
\begin{equation} \varrho=\sum_{i=1}^kp_i \proj{e_i, f_i}, \ \ \ \sum_i p_i=1,
\label{separable}
\end{equation}
where $\ket{e_i,f_i}$ stands here for a normalized vector $\ket{e_i}\otimes\ket{f_i} \in {\cal H}_{A} \otimes {\cal H}_{B}$.
 In finite dimensional cases, the ones we will be concerned here, the approximation part
is not necessary, as for any separable state one can always find 
a set $\{\ket{e_i,f_i}\}$ of product vectors  for which $k\leq
{\rm dim} ({\cal H}_{AB})^2$ in the above formula \cite{tran}.

Several necessary conditions for separability are known: 
Werner derived a condition based on the analysis of 
local hidden variables models and mean value of the, so called,
flipping operator
\cite{Werner}; the Horodeckis proposed a necessary criterion based 
on  $\alpha$-entropy inequalities
\cite{alfa}. 
Peres  demonstrated  that the partial transpose 
$\varrho^{T_A}$ of the matrix $\varrho$, 
defined as
$\langle m ,\mu|\varrho^{T_A}| n ,\nu\rangle
= \langle n ,\mu|\varrho| m ,\nu\rangle$
for any fixed product basis $|n ,\nu \rangle \equiv 
|e_n \rangle_{A} \otimes |e_\nu \rangle_{B}$,
must be still a legitimate density matrix if $\varrho$ is separable \cite{Peres}.
This operationally friendly, necessary condition, called  positive partial 
transpose (PPT) condition, turned out to be very strong.

Soon after Peres result, a general connection between 
positive map theory and separability was established in \cite{sep}, where 
necessary and sufficient separability conditions were derived in terms of positive maps.
In particular it implied that for systems of low dimensions ($M N\le 6$)
the PPT condition is also sufficient for separability. 
It was also shown that this is not the case
for systems of higher dimensions ($M N> 6$).
Later on explicit counterexamples of entangled
states with PPT property were provided
 by means of another separability criterion,
based on the analysis of the range of the density matrix \cite{tran}
(cf. \cite{Woronowicz}).
It was then shown that they represent 
bound entanglement \cite{bound}.
Let us note that on mathematical grounds there were 
examples, provided earlier \cite{Choi},
of elements  of positive matrices cones
which can be treated as prototypes of 
PPT entangled states.

Sufficient conditions for separability are also known. We remark that the results of \cite{Zyc} readily imply that any state close enough to the completely random state $\pi$ is separable. Thus, as quantified in \cite{Rob}, any mixture $\tilde{\varrho} = (\!1\!-\!p)\varrho + p \pi$ in a $M\times N$ system is separable if $p \geq (1+2/MN)^{-1}$ or, in other words, as we wish to make explicit here, a full rank mixed state is separable provided its smallest eigenvalue be greater than or equal to $(2+MN)^{-1}$.

 On the other hand the analysis of the range of the density matrices,
first applied in the separability criterion \cite{tran},
led to an algorithm for the optimal decomposition
of mixed states into a separable and an inseparable part \cite{M&A}, and
to a systematic method of constructing
examples of PPT entangled states and peculiar positive maps \cite{UPB,terhal}.
Also, the technique of diminishing the rank of a PPT density matrix
by subtraction of selected product vectors, which was worked out in \cite{STV},
 turned out to be very useful. This and other techniques have allowed quite recently to study operational necessary and sufficient separability conditions for states of a $2 \times N$ system \cite{MCK}. In particular it has been shown that:

($i$) all PPT states of rank smaller than $N$ are separable;

($ii$) the separability of generic states such that $r(\varrho)+r(\varrho^{T_A}) 
\le 3N$ reduces to analyzing the roots of some complex polynomials (a constructive separability criterion was derived, thus providing also the decomposition of such separable states into pure product states);

($iii$) states invariant under  partial transpose, and those 
that are not ``very different'' from their partial transpose are 
necessarily separable.

This paper can be considered an extension and generalization of Ref. \cite{MCK}. 
The results ($i$) and ($ii$) obtained there for $2\times N$ systems are here generalized non-trivially to the case of $M\times N$ systems ($M \leq N$). We show, namely, that 

\begin{itemize}

\item any state $\varrho$ supported on $M \times N$ ($M \leq N$) and with rank $r(\varrho)\leq N$ is separable iff its partial transpose is positive;

\item separability of generic PPT density matrices with $r(\varrho)+r(\varrho^{T_A}) \le 2MN-M-N+2$ reduces to solving a system of coupled polynomial equations.

\end{itemize}

In both cases a pure product state decomposition for separable states is obtained.
 
 Throughout this paper we make use of the following definition: we say that a state $\rho$ acting on $M \times N$ is {\em supported} on $M\times N$ if this is the smallest product Hilbert space on which $\rho$ can act. Let us introduce the local ranks $r(\varrho_A)$ and $r(\varrho_B)$, where $\varrho_{A,B} \equiv $ Tr$_{A,B} \varrho$ are the reduced density operators.
 It immediately follows from the first of the above results that there is no PPT bound entanglement of rank 3. Indeed, a rank $3$ state $\rho$ either has at least one of the local ranks $r(\rho_A)$ and $r(\rho_B)$ greater than $3$, and in this case is distillable \cite{rank2} (i.e., $\rho^{T_A}$ is not positive), or else can be supported on a $MN \leq 6$ or on a $3 \times 3$ system, and thus is separable. This implies in particular that the bound entangled states constructed with the UPB method
\cite{UPB} and those based on the chess-board structure
of eigenvectors \cite{Dagmar} are {\it optimal} with respect to
their ranks.

 For our second main result,  concerned with those PPT density
matrices for which the sum of ranks satisfies $r(\varrho)+r(\varrho^{T_A})
\le 2MN-M-N+2$, we identify the eligible product vectors (that is, those that can appear in decomposition (\ref{separable}) if $\rho$ is separable) with the solutions of a system of coupled polynomial equations. We analyze these equations, which are arguably expected to have only a finite number of solutions. For this case we present a constructive (i.e. leading to a product state decomposition) method to check separability. Also for the same case we discuss an alternative, constructive method to check separability numerically. These checks represent a necessary and sufficient condition for separability.

 We wish to remark the importance of having separability conditions for low rank density matrices, especially in relation to unsolved problems concerning the nature of bound entanglement (BE). Notice that such conditions are of great value when trying to construct states with BE. Among the open questions we encounter the existence of BE having a non-positive partial transpose NPT (see \cite{evidence}). Also, whether a finite or a vanishing amount of free entanglement is required to asymptotically create bound entangled states. There are in addition several conjectures concerning bound entanglement (see \cite{bound,UPB,aktyw,bechan,Vedral}) among them the ones connected to capacities of quantum channels and bound entanglement assisted distillation. Finally, we have been recently able to establish a general connection between low rank bound entangled states and positive maps. This connection allows for a systematic construction of independent linear maps in arbitrary dimensions, including $2\times N$, where the procedures based on unextendible product bases do not work \cite{terhal}. The discussion of this connection will be presented elsewhere.

 This paper is organized as follows: we start by generalizing some needed results of \cite{MCK} related to diminishing the rank of $\rho$ by subtracting projectors on product vectors; in Section III we present our theorem about the separability of states with rank $ \leq N$;  in Section IV the necessary and sufficient separability conditions for generic matrices with $r(\varrho)+r(\varrho^{T_A})\le 2MN-M-N+2$ are formulated, and discussed in the context of $3\times 3$ systems; finally, Section V contains our conclusions and acknowledgments.

\section{Diminishing of the rank - generalizations}

Before we turn to the main results of this paper 
we need to generalize some of those presented in \cite{MCK}.
 
Consider a state $\varrho$ of a $M\times N$ system satisfying $\varrho^{T_{A}}\geq 0$.
Throughout this paper $K(X), R(X), k(X)$, and $r(X)$ denote the kernel,
the range, the dimension of the kernel, and the rank of the operator $X$,
respectively. By $\{\ket{a_i}\}_{i=1}^M$ and $\{\ket{b_i}\}_{i=1}^N$ we will denote orthonormal basis in ${\cal H}_A$ and ${\cal H}_B$, and by $| e^{*}\rangle$ we will denote the complex conjugated
vector of $| e\rangle$ in the orthonormal basis $|1\rangle_{A}, \cdots, |M\rangle_{A}$ in
which we perform the partial transposition; that is, if $\ket{e} = \sum_{i=1}^M \alpha_i \ket{i}$ then $\ket{e^{*}} = \sum_{i=1}^M \alpha_i^{*} \ket{i}$.

For the time being we do not require $M\leq N$. The following Lemma is a generalization of Lemma 6 of Ref. \cite{MCK} proved there for $M=2$:

{\bf Lemma 1.-} If $\exists$ $\ket{f} \in {\cal C}^{N}$ such that 
$|a_i , f\rangle \in K (\varrho) $ for $i=1, ..., M-1$, 
then either
(i) $|a_M , f\rangle \in K (\varrho) $
or (ii)
\begin{eqnarray}
&&\varrho |a_M , f\rangle=|a_M , g\rangle \nonumber \\
&&\varrho^{T_{A}} |{a_M}^* , f\rangle=| {a_M}^* , g\rangle
\label{ii}
\end{eqnarray}
for some $|g\rangle \in {\cal C}^{N}$.

{\it Proof.-}
From the assumptions we have immediately $\varrho^{T_A}|{a_i}^* , f\rangle=0
$ ($i=1, ..., M-1$). In particular  $\forall |h\rangle\in {\cal C}^{N}$
we have  $ \langle {a_M}^*, h |\varrho^{T_A}|{a_i}^* , f\rangle=0$
or, equivalently, $\langle a_i, h |\varrho|a_M , f\rangle=0$.
Since $|h\rangle$ is arbitrary we have either statement (i)
or $\varrho|a_M , f\rangle= |a_M , g\rangle$ for some $ |g\rangle \neq 0$.
The second case needs further analysis.
In a similar way we can prove that either
$\varrho^{T_A}|{a_M}^* , f\rangle=0$ (which is still equivalent to
the statement (i))
or $\varrho ^{T_A}|{a_M}^*, f\rangle=|{a_M}^*, g'\rangle$
for some $|g'\rangle\neq 0$.
It remains to prove that $|g'\rangle=|g\rangle$.
Indeed $|g'\rangle =\langle{a_M}^*|\varrho ^{T_A}|{a_M}^*, f\rangle=
\langle a_M|\varrho|a_M , f\rangle= |g\rangle$.

The second lemma below is also  
a generalization of the results from Ref. \cite{MCK}: 

{\bf Lemma 2.-} If $\varrho$ satisfies the assumptions of 
Lemma 1,  and the possibility (ii) of Lemma 1 holds, 
then 
\begin{equation}
\varrho_1=\varrho - \lambda |a_M, g \rangle \langle a_M, g |,
\label{ro1}
\end{equation}
where $\lambda \equiv \langle a_M, g| \varrho^{-1} |a_M, g \rangle$ and

\vspace{4mm}

\noindent (i) $\varrho_1$ is a PPT state with $r(\varrho_1) = r(\varrho) - 1$ and $r(\varrho_1^{T_A}) = r(\varrho^{T_A}) - 1$.
\vspace{4mm}

\noindent (ii) $\varrho_1$ is supported either on a $(M\!-\!1) \times (N\!-\!1)$ or on a $M \times (N\!-\!1)$. 
\vspace{4mm}

\noindent (iii) $\varrho_1$ is separable iff $\varrho$ is separable.
\vspace{4mm}

{\it Proof .-} Following Corollary 1 and Lemma 2 from Ref. \cite{MCK}, we observe that

\begin{equation}
\varrho_1=\varrho -\frac{|a_M, g \rangle \langle a_M, g | }{
\langle a_M, g| \varrho^{-1} |a_M, g \rangle}
\end{equation}
is positive, and that $r(\varrho_1) = r(\varrho) - 1$. Then (i) follows from taking into account that since 
\begin{eqnarray}
\lambda^{\!-\!\!1} &=& \langle a_M, g| \varrho^{-1} |a_M, g \rangle = \nonumber\\
\langle g|f \rangle&=&\langle {a_M}^*, g| {\varrho^{T_A}}^{-1} |{a_M}^*, g \rangle,
\label{lambda}
\end{eqnarray}
we have that 
\begin{equation}
{\varrho_1}^{T_A}=\varrho^{T_A}
-\frac{|{a_M}^*, g \rangle \langle {a_M}^*, g | }{
\langle {a_M}^*, g| {\varrho^{T_A}}^{-1} |{a_M}^*, g \rangle},
\label{ro1a}
\end{equation}
so that also $\varrho_1^{T_A}$ is positive and $r(\varrho_1^{T_A}) = r(\varrho^{T_A}) - 1$. From assumptions on the kernel of $\varrho$ it follows that
all vectors $\{|a_i, f\rangle \}_{i=1}^{M}$ belong 
to the kernel of $\varrho_1$, hence  $\varrho_1$
can be embedded into a $M\times(N-1)$ space; on the other hand since $\varrho_{1,A} = \varrho_A - \lambda\braket{g}{g}\proj{a_M}$, $r(\varrho_{1,A})$ must be either $M=r(\varrho_A)$ or $M-1$,  which finishes the proof of (ii). In order to prove (iii) let us assume that $\varrho$ is separable and let us show that also $\varrho_1$ is so (if $\varrho_1$ is separable then  obviously $\varrho$ is also separable). Since $\varrho \ket{a_i, f} = 0$ ($i=1, ..., M-1$), we can always write  
\begin{equation}
\varrho = \sum \proj{e_i, f_i} + \proj{a_M}\otimes\eta,
\end{equation}
where $\braket{f}{f_i} = 0$ and $\eta$ is a positive operator acting on ${\cal C}^N$. If we impose $\ket{a_M,g} = \varrho \ket{a_M, f}$ we obtain $\ket{g} = \eta\ket{f}$, and therefore $\ket{g}\in R(\eta)$. We can write
\begin{equation}
\varrho_1 = \sum \proj{e_i, f_i} + \proj{a_M}\otimes(\eta-\lambda\proj{g}),
\end{equation}
so that if we show that the operator $(\eta - \lambda\proj{g})\geq 0$ then we have that $\rho_1$ is separable. Using (\ref{lambda}) we have that such an operator is
\begin{equation}
\eta - \frac{1}{\braket{g}{f}}\proj{g} = \eta - \frac{1}{\bra{g}\eta^{-1}\ket{g}}\proj{g},
\end{equation}
and that therefore it is positive (cf. Lemma 1 in  \cite{MCK}).
$\Box$

\section{All rank $N$ PPT states supported on a $M \times N$ system ($M\leq N$) are separable}

In this section we generalize the following theorem
proved in Ref. \cite{MCK}: 

{\bf Theorem.-}[Theorem 1 of Ref. \cite{MCK}]

Let $\varrho$ be a PPT state of rank $N$ supported on a $2 \times N$ space.
Then $\varrho$ is separable and can be written as
\begin{equation}
\varrho=\sum_{i=1}^{N} |e_i, f_i \rangle \langle e_i, f_i|
\label{conv}
\end{equation}   
 with all $\{ \ket{f_i} \}$ linearly independent.

We will express a density matrix in terms of its reduced operators $\bra{i_A}\varrho\ket{j_A}$ acting on ${\cal H}_B$. For instance, we will write (\ref{conv}) as 
\begin{eqnarray}
\varrho=\left[ \begin{array}{cc}
         \tilde{A} & \tilde{B}^{\dagger} \\
         \tilde{B} & \tilde{C} \\
       \end{array}
      \right ],
\label{canon'}
\end{eqnarray}
where $\tilde{A} \equiv \bra{1}\rho\ket{1}\geq 0$, $\tilde{C} \equiv \bra{2}\rho\ket{2}\geq 0$ and $\tilde{B} \equiv \bra{2}\rho\ket{1}$. More generally, a $M \times N$ density matrix will be expressed as
\begin{eqnarray}
\varrho=\left[ \begin{array}{cccc}
         E_{11} & E_{12} & ... & E_{1M} \\
         E_{12}^\dagger & E_{22} & ... & ... \\
          ...& ...  & ...& ...  \\
          E_{1M}^{\dagger} & ... & ...  & E_{MM} 
       \end{array}
      \right ],   
\label{canoni}
\end{eqnarray}
where now $E_{ij} \equiv \bra{i_A}\varrho\ket{j_A}$. We start by using the previous Theorem to prove:

{\bf Lemma 3.-} Let $\varrho$ be a rank $N$ PPT state supported on a $2\times N$ space.
Then after a reversible local filtering operation
\footnote{A reversible transformation is a transformation that can be reversed with nonzero probability. A local filtering in Bob's side $I\otimes V$ is then reversible iff the operator $V$ can be inverted, i.e. iff $V^{-1}$ exists.}
the state is proportional to 
the matrix (called hereafter $2 \times N$ {\em canonical form}):
\begin{eqnarray}
\Sigma\equiv \left[ \begin{array}{cc}
         B^{\dagger}\!B & B^{\dagger} \\
         B & I \\
       \end{array}
      \right ] = [B~~I]^{\dagger}[B~~I],
\label{canon} 
\end{eqnarray}
with $B$ normal, i.e. $[B,B^\dagger]=0$.
 
{\it Proof .-} We write the density matrix (\ref{conv}) in the 
form (\ref{canon'}).
Because there is only a finite number of $\ket{e_i}$ in (\ref{conv}) we can always find a vector $\ket{a}$ such that $\langle a|e_i \rangle \neq 0$
for all $i$.  Let this $|a\rangle$ be the second element of the orthonormal basis in Alice's space, i.e. $\ket{2}= \ket{a}$. The matrix $$\tilde C=\langle a|\varrho|a\rangle=
\sum_{i=1}^{N}|\langle a |e_i\rangle|^2| f_i \rangle \langle f_i|$$ 
has then maximal 
rank $N$, since the $\ket{f_i}$ are linearly independent. Taking the local filter $V=(\sqrt{\tilde{C}})^{-1}$ on Bob's side (this corresponds 
to sandwitching the state between $I \otimes V$
and $I \otimes V^{\dagger}$) we obtain 
\begin{eqnarray}
\tilde{\varrho}=\left[ \begin{array}{cc}
         A \ B^{\dagger} \\
         B \ I \\
       \end{array}
      \right ],
\label{canon1}
\end{eqnarray}
which is still positive and PPT (because any local operation
preserves the PPT property \cite{bound}).
We can write
\begin{equation}
\tilde{\varrho}=\Sigma + {\rm diag}[\Delta,0],
\end{equation}
where the positive matrix $\Sigma$ from the expression (\ref{canon}) 
has rank $r(\Sigma)=N$ as its kernel $K(\Sigma)$ has 
at least dimension $N$ containing all 
 vectors of the type 
\begin{equation}
|\phi_f \rangle = |1 \rangle |f\rangle +|2 \rangle |-Bf \rangle, 
\label{kernel}
\end{equation}
 while its range
has at least dimension $N$ due to the identity entry on the 
diagonal. 
Notice that diag$[\Delta, 0]$ is also positive, because positivity of $\tilde{\varrho}$ implies that $\Delta=A- B^{\dagger}B \geq 0$ \cite{truco}. Now, since in addition $r(\tilde{\rho}) = r(\Sigma)$, we also have that $R(\tilde{\rho}) = R(\Sigma) \supseteq R($diag$[\Delta, 0])$, that is $K($diag$[\Delta, 0]) \supseteq K(\Sigma)$. But $K(\Sigma)$ is spanned by the states (\ref{kernel}), for which then $\bra{\phi_f}$diag$[\Delta, 0] \ket{\phi_f}=0$, which finally implies $\Delta\ket{f} = 0~~\forall \ket{f}$.
This ends the proof of 
the fact that $\Delta=0$,  or in another words that $A=B^{\dagger}B$.
This proves therefore the canonical form 
(\ref{canon}), but not yet  the normality of $B$.
The latter property can be simply proven from the positivity of $\tilde{\varrho}^{T_A}$, which implies that $BB^\dagger - B^\dagger B\geq 0$ \cite{truco}. The latter (positive) operator 
has at the same time null trace and therefore it must vanish. 
Thus $B$ is normal as stated.$\Box$

Let us prove now the generalization of Lemma 3 to the case of $3\times N$ systems ($N\geq 3$), and then to the $M\times N$ case, where $M\leq N$ from now on.

{\bf Lemma 4.-} Let $\varrho$ be a PPT state of rank  $N$ in a $3\times N$ 
space. Let the  reduced state $\varrho_B$
and  the entry $E_{33}$ in some local basis
have  also the same rank $N$.
Then $\varrho$ can be transformed using some 
reversible local transformation to the canonical form:
\begin{equation}
\varrho \sim [C,B,I]^\dagger[C,B,I]
\label{forma}
\end{equation}
where $C,B$ are normal and $[B,C^\dagger]=[B,C]=0$.

Note that in Lemma 4, in contrast to Lemma 3, 
 we assume that in some basis $r(E_{33})=N$. 
Later on, in Theorem 1, we will prove that this assumption is always satisfied.

{\it Proof .-} In order to obtain the  identity matrix $I$ at the 
diagonal we use an analogous reversible
local filter to the one used in the proof of Lemma 3.
After that we readily obtain the form

\begin{eqnarray}
\varrho \sim \tilde{\varrho}=\left[ \begin{array}{ccc}
          C^\dagger\! C & D^{\dagger}  & C^\dagger  \\
         D  &  B^\dagger\! B  & B^\dagger \\
        C & B  & I\\
       \end{array}
      \right ].
\label{canon2a}
\end{eqnarray}
with both $B$ and $C$ normal and some unknown $D$. Indeed, expression (\ref{canon2a})  as well as
the normality of $B$ and $C$ follow from the fact that 
after a local projection by projectors $P_k \otimes I \equiv (|k \rangle \langle k| +
|3\rangle\langle 3|) \otimes I$, $k=1,2$  
we get a $2\times N$ state satisfying the assumptions of Lemma 3. 

Now, notice that $\langle \Psi_f | \tilde{\varrho} |\Psi_f \rangle=0$
for $|\Psi_f \rangle\equiv |2 \rangle |f\rangle -|3\rangle|Bf \rangle$.
Since $\tilde{\varrho}\geq 0$ we have that
\begin{equation}
0 = \tilde{\varrho} |\Psi_f \rangle=|1\rangle|D^{\dagger}f - C^\dagger B f\rangle,
\end{equation}
which, as $f$ is arbitrary, leads to
$D^\dagger=C^\dagger B$. Thus formula (\ref{forma}) holds.
Finally e shall use the latter as well as normality of $B$ and $C$
to prove that $[B,C^\dagger]=[B,C]=0$.
We have
\begin{eqnarray}
\varrho^{T_A} \sim \tilde{\varrho}^{T_A}=\left[ \begin{array}{ccc}
          C^\dagger C & B^\dagger C & C  \\
          C^\dagger B & B^\dagger B & B \\
          C^\dagger   & B^\dagger   & I\\
       \end{array}
      \right ],
\label{canon2}
\end{eqnarray}
and we can check that for any $|f\rangle \in {\cal C}^N$ and for $|\Phi_f \rangle \equiv |2\rangle|f\rangle-|3\rangle |B^{\dagger} f\rangle$, $\langle \Phi_f | \tilde{\varrho}^{T_{A}} |\Phi_f \rangle=0$.
As $\varrho$ is PPT this implies that
\begin{equation}
 \varrho^{T_A} |\Phi_f \rangle= |1 \rangle|[B^\dagger,C]f \rangle
\label{zero}
\end{equation}
must vanish. Since the above equation holds for arbitrary
$| f \rangle $ we have immediately that
$[B,C^\dagger]=[C,B^\dagger]^\dagger=0$ Normality of $B$ and of $C^{\dagger}$ implies that these operators can be decomposed as a complex linear combination of projectors into eigenvectors. That they commute means that they actually have the same eigenvectors, and thus so do $B$ and $C$, i.e. $[B,C]=0$.$\Box$

{\bf Lemma 5.-} 
Any PPT state supported on a $M \times N$ space ($M\leq N$) satisfying that  
(i) $r(\varrho)=N$,  (ii) in some product basis 
 $r(E_{ii})=N$ for some $i$, 
can be transformed after a reversible local transformation
to the canonical form:
\begin{equation}
\varrho \sim Z^\dagger Z=[C_1,...,C_{M-1},I]^\dagger[C_1,...,C_{M-1},I]
\label{formaca}
\end{equation}
with $[C_i,C_j^\dagger]=[C_i,C_j]=0$, $i,j=1, ..., M-1$.

{\it Proof .-} It follows easily from 
the application of  Lemmas 3 and 4. 
In particular one has to use the local projections
$P \otimes I =(|k \rangle \langle k| +
|M\rangle\langle M|) \otimes I$, $1\leq k < M$,
$P' \otimes I =(|m \rangle \langle m| +
|m' \rangle \langle m'| +
|M\rangle\langle M|) \otimes I$, $1\leq m < m' < M$.$\Box$

As an immediate consequence we have

{\bf Lemma 6 .-} Any PPT state supported on a $M \times N$ space ($M\leq N$) satisfying that  
(i) $r(\varrho)=N$,  (ii) in some product basis 
 $r(E_{ii})=N$ for  some $i$, is separable and can be expressed as
\begin{equation}
\tilde{\varrho}=\sum_{i=1}^N |e_i, f_i \rangle \langle e_i, f_i|,
\label{lemma6}
\end{equation}
where the $\{\ket{e_i}\}$ are possibly unnormalized and the $\{\ket{b_i}\}$ are linearly independent.

{\it Proof .-} We make use of Lemma 5.
It is easy to see that the matrix $Z^\dagger Z$ 
has nonzero eigenvectors of the form $|a_i, b_i \rangle$.
Here $|b_i \rangle$ is
the $i-$th common eigenvector 
of all operators $C_j$, $C_k^\dagger$ 
while $|e_i\rangle^{\dagger}=
[c_i^{(1)}, . . . , c_i^{(M-1)}, 1]$
is a row of all $i-$th eigenvalues of matrices $C_1, ..., C_{M-1}, I$.
Thus, after some reversible local transformation the state $\rho$ becomes:
\begin{equation}
\tilde{\varrho}=\sum_{i=1}^N |e_i, b_i \rangle \langle e_i, b_i|,
\end{equation}
where the $\{ |b_i\rangle \}$ are orthonormal. Reversing the previous local filtering we obtain (\ref{lemma6}) $\Box$.

{\bf Remark .-} The above procedure gives a constructive 
algorithm to decompose any state which satisfies the assumptions
of the Lemma.

The main disadvantage of the above results is that all of them contain 
assumptions about $r(E_{ii})=N$ for some $i$ and for some product basis, 
which as we have mentioned, are not necessary. Our main theorem is free 
of that assumption (i.e., it shows that such $E_{ii}$ always exists). To prove it we have to use induction with respect to $M+N=K$ and use the previous Lemmas. We consider only $r(\varrho) = N$, as a PPT state supported on $M\times N$ cannot have smaller rank. Indeed, since $r(\varrho_B) = N$, if $r(\varrho) < N$ then $\varrho$ is distillable, which implies that $\varrho^{T_A}$ is not positive \cite{rank2}.

{\bf Theorem 1.-} All rank $N$ PPT states $\varrho$ supported on $M\times N$ are separable.

{\it Proof.-} We will prove that in some product basis we have $r(E_{ii})=N$ for some $i$. Separability of $\varrho$ will follow from the previous Lemmas.
Let us observe that the Theorem and the latter fact are true for $M=2$ and arbitrary $N\ge 2$.
In particular they are true for $M+N=K=4$ and 5. Let us assume that they hold for $M+N\le K$. We shall now demonstrate that 
they also hold for $M+N=K+1$.  

To this aim let us consider the case of $\varrho$ supported on a $(M +1)\times N$ space, with $M+1\le N$, 
$r(\varrho_A)=M+1$, $r(\varrho_B)=N$.
and $M+N=K$. In an orthonormal, product basis representation 
the state $\varrho$
has  the form of a $(M+1) \times (M+1)$ matrix with $N\times N$ entries
\begin{eqnarray}
\varrho=\left[ \begin{array}{cccc}
         E_{11} & E_{12} & ... & E_{1,M+1} \\
         E_{12}^\dagger & E_{22} & ... & ... \\
          ...& ...  & ...& ...  \\
          E_{1,M+1}^{\dagger} & ... & ...  & E_{M+1,M+1}
       \end{array}
      \right ].
\label{canonm+1}
\end{eqnarray}
Let us consider the following 
$M \times M$ submatrix of 
$\varrho$ 
\begin{eqnarray}
W(\varrho)\equiv W=\left[ \begin{array}{cccc}
         E_{22} & E_{23} & ... & E_{2,M+1} \\
         E_{23}^\dagger & E_{22} & ... & ... \\
          ...& ...  & ...& ...  \\
          E_{2,M+1}^{\dagger} & ... & ...  & E_{M+1,M+1}
       \end{array}
      \right ],
\label{obciecie}
\end{eqnarray}
resulting after removing the first row and the first column from
the representation (\ref{canonm+1}). As the latter action can be achieved 
by a local projection 
on Alice's side,  $W$ is an unnormalized PPT state acting in $M \times N$.
For Bob's reduced matrix $W_B=Tr_B(W)$,
we shall consider two alternative possibilities: 
(i) $r(W_B)=N$, (ii) $r(W_B)<N$.

In case (i) we must have $r(W)=N$ as otherwise
we would have that the global rank is less than one of the local ranks, 
resulting in distillability of $\varrho$, 
{\it ergo} in  violation of the PPT condition \cite{bound,rank2}.
But it means that $W$
is a PPT state supported on $M \times N$ with global and 
local rank equal to $N$. According to the induction assumption, it is 
thus separable 
and for some product basis has an entry  $E_{ii}$ for some $i=2,\ldots,M+1$ 
with the 
 rank $N$. But then $\varrho$ has an entry $E_{ii}$
with rank $N$ in the same product basis, and 
from Lemmas 5 and 6 it follows immediately that 
it is separable.

Consider now case (ii). 
If $W$ has $r(W_B)<N$,
then obviously there exist a sequence 
of product vectors 
$|a_i , f\rangle \in K (W) $,
$i=2, ..., M+1$. 
It is immediate to check that they must
belong to kernel of $\varrho$. That means that the assumptions 
of the Lemma 1 are fulfilled. The possibility (i) of this Lemma 
cannot hold because otherwise one could embed $\varrho$ into 
$(M+1)\times(N-1)$ space, and $r(\varrho_B)$ would be $N-1$ instead of $N$. 
The possibility (ii) of Lemma 1 means that $\varrho$ can be written 
in the form (cf. Lemma 2)
\begin{equation} 
\varrho=\varrho'+\lambda |1,g\rangle \langle 1,g|
\label{redukcja}
\end{equation}
where $\varrho'$ is a rank $N-1$ PPT state supported either on a $(M+1) \times (N-1)$ subspace or on a $M \times (N-1)$ subspace, $\lambda^{-1} \equiv \bra{1,g}\rho^{-1}\ket{1,g}$ 
and  $|1,g\rangle \langle 1,g|$ is an unnormalized
 projector onto a product state
such that $\varrho^{-1}|1,g\rangle$ is orthogonal to $R(\varrho')$.

At the same time it must hold that $r(\varrho'_B)=N-1$, since i) 
Bob's space has now only
$N-1$ dimensions, ii) $r(\varrho'_B)$ cannot be smaller than $N-1$, since $N=r(\varrho_B = \varrho'_B + \proj{g})$ and $\proj{g}$ can increase at most in one unit the rank of $\varrho'_B$.
All that means that the matrix $\varrho'$ fulfills the 
induction assumption as $(M+1)+(N-1)=K$ (or $M+(N-1)=K-1$ ) and $r(\varrho')=r(\varrho'_B)$,
{\it ergo} it is separable and has in some product basis $\ket{a_i, b_j}$ the 
entry $E'_{ii}=\langle a_i|\varrho'|a_i\rangle$ with rank $N-1$. Lemma 6 implies then that $\varrho$ ($=\varrho' + \lambda \proj{1,g}$) can be decomposed into
\begin{equation} 
\sum_{i=1}^{N-1} |e_i, f_i \rangle \langle e_i, f_i| + \lambda \proj{1,g},
\end{equation}
where $\ket{g}$ is linearly independent from the set of (also linearly independent) vectors $\ket{b_i}$. Since there is only a finite number of projectors in the decomposition above, we can always find a vector $\ket{a}$ in Alice's space such that $\braket{e_i}{a}\neq 0 \neq \braket{1}{e}$. Including such a vector in a product basis to express $\varrho$ we will obtain the wished rank $N$ element $\bra{a} \varrho \ket{a}$.  
This completes the proof of the induction step, and  by induction 
completes thus the proof of the theorem. $\Box$

\section{Separability criteria 
for rank$(\varrho)+$rank$(\varrho^{T_A})\le 2MN-M-N+2$}

In this section we generalize the results obtained for $2\times N$ systems
in Ref. \cite{MCK}. The idea is that a PPT density operator $\varrho$ with  
$r(\varrho)+r(\varrho^{T_A})\le 2MN-M-N+2$ may have a finite 
number of product vectors $|e_i,f_i\rangle$ in its range, such that $|e_i^*,f_i\rangle
\in R(\varrho^{T_A})$. These product vectors are the only possible candidates to appear in decomposition (\ref{separable}) \cite{tran}. Finding them requires solving a system of polynomial equations. First we show how to solve these equations in a {\em generic} case (namely when the coefficients of such equations do not happen to satisfy a large series of conditions, which amounts to having only a finite number of solutions), and once all the product states $\{\ket{e_i,f_i}\}_{i=0}^{L<\infty}$ have been obtained, we present an algorithmic method to check whether $\rho$ is separable. This is done in a finite number of computational steps, and thus solves operationally the problem of separability for states with $r(\varrho)+r(\varrho^{T_A})\le 2MN-M-N+2$ and finite $L$. 

\subsection{Eligible product vectors.}

Let the linearly independent 
vectors $|K_i\rangle$, $|\tilde K_i\rangle$  form a basis in the kernel
of $\varrho$ and in the kernel of $\varrho^{T_A}$, respectively:
\begin{eqnarray}
K(\varrho)&=& \mbox{span} \{  |K_i\rangle, i=1, ..., k(\varrho) \} \\
K(\varrho^{T_A})&=&\mbox{span} \{  |\tilde{K}_i\rangle, i=1, ...,
k(\varrho^{T_A}) \}
\end{eqnarray}
We consider here the case when $k(\varrho)+k(\varrho^{T_A})\ge M+N-2$. 
We can always expand $|K_i\rangle$ and $|\tilde K_i\rangle$ in an orthonormal
basis in Alice's space
\begin{eqnarray}
|K_i\rangle& =&\sum_{m=0}^{M}|m,k_i^{m}\rangle,\label{ozn0}\\
|\tilde{K}_i\rangle&=&\sum_{m=0}^{M}|m,\tilde{k}_i^{m}\rangle.
\label{ozn}
\end{eqnarray}
A product vector $|e,f\rangle$  
belonging to the range $R(\varrho)$ must be orthogonal to 
all $|K_i\rangle$; simultaneously, 
if its partial complex conjugation belongs 
to $R(\varrho^{T_A})$, $|e^*,f\rangle$ must be orthogonal to 
all $|\tilde K_i\rangle$. 
Thus the eligible product vectors are the solutions of 
$k(\varrho)+k(\varrho^{T_A})$ equations, namely
\begin{eqnarray}
&&\langle K_i|e,f\rangle =0\nonumber, i=1, ..., k(\varrho),\ \\
&&\langle \tilde{K}_i|e^*,f\rangle=0,  \ i=1, ..., k(\varrho^{T_A}).
\label{conditions}
\end{eqnarray}
Let us now expand $|e \rangle$ in the above formula as:
\begin{equation}
|e \rangle = \left[ \begin{array}{c} \alpha_{1} \\ \vdots\\ \alpha_{M} \end{array} \right].
\end{equation}

We restrict ourselves to $\alpha_1=1$. The reason is that we expect to find only a {\it finite} number $L$ of inequivalent vectors $|e_i,f_i\rangle$ that fulfill the requirements. A generic choice of an orthonormal basis $\{\ket{a_i}\}$ in Alice's space will imply that $\langle 1|a_i\rangle\ne 0$ for all $i=1,\ldots,L$. In this basis $\alpha_1$ can be set equal to 1. 

Equations (\ref{conditions}) can be rewritten as follows:
\begin{equation}
A(\alpha_1, ..., \alpha_{M};\alpha_1^*, ..., \alpha_{M}^*)|f\rangle=0,
\label{conditions2}
\end{equation} 
where the $[k(\varrho)+k(\varrho^{T_B})] \times N$ matrix $A$ is 
defined as follows: 
\begin{eqnarray}
&&A(\alpha_1, ..., \alpha_{M}; \alpha_1^*, ...,
\alpha_{M}^*)\equiv
\left[ \begin{array}{c}
         \sum_{m=1}^M \alpha_m \langle k_1^m |   \\
           ... \\
           \sum_{i=1}^M  \alpha_m \langle k_{k(\varrho)}^m|\\
          \sum_{i=1}^M  \alpha_m^* \langle \tilde{k}_1^m|\\
          ...  \\
        \sum_{i=1}^M\alpha_m^* \langle \tilde{k}_{k(\varrho^{T_A})}^m|
       \end{array}
      \right ] \nonumber \\ 
&&\equiv\left[ \begin{array}{c}
      D_{k(\varrho) \times N}(\alpha)  \\
      \tilde{D}_{k(\varrho^{T_B}) \times N}(\alpha^*)             
      \end{array}
      \right ].
\label{matirx}
\end{eqnarray}

If (\ref{conditions2}) holds for some $\ket{f}\neq 0$ and $\ket{e}\neq 0$, this means that for the corresponding set of $\alpha$'s the rank of $A$ is smaller than $N$. Therefore, in order to identify eligible product vectors we have to require that at most $N-1$ rows of $A$ be linearly independent vectors. In what follows we restrict ourselves to the limiting case $k(\varrho)+k(\varrho^{T_A})= M+N-2$, the others containing more restrictions and consequently less solutions than this.

Let us then take $N-1$ rows of $A$, say the first ones, and let us require that each of the remaining $M-1$ rows be linearly dependent of them. Recall that we can use the $M-1$ variables $\alpha_2, \cdots, \alpha_M$ in order to achieve this. Then, parameter counting strongly suggests that we need to fix all the $M-1$ $\alpha$'s in order to make $A$ have rank smaller than $N$, this corresponding to a zero measure set of points in the $\alpha$-space $[\alpha_1\!\!=\!\!1, \alpha_2, \cdots, \alpha_M]$. We will in addition relate the number of solutions to the roots of complex polynomials, which under {\em generic} conditions have only a finite number of roots.  Numerical experience acquired for the $2\times N$ case further supports the expectation that the number of solutions {\em be} typically finite. 

\subsection{Generic Polynomials.}

 Let us discuss a bit further sufficient conditions for the existence of a finite set of solutions, while presenting a systematic method to find them once the conditions are fulfilled. This method also works for $k(\varrho)+k(\varrho^{T_A}) < M+N-2$ by just adding more equations. 

Matrix $A$ will have at most rank $N-1$ after requiring that all its rank $N$ minors vanish. At risk of finally finding more solutions than just those of equations (\ref{conditions2}), we can impose that only $M-1$ of these minors vanish. The reason for doing so is that this will already allow us to prove that only a finite number of product vectors fulfill (\ref{conditions}) under some {\em generic} circumstances. Thus we consider the determinant of $N\times N$ submatrices of $A$ formed by taking its first $N-1$ rows and then also one of the $M-1$ remaining ones. We shall denote these minors by $F_i(z_1, \ldots, z_{2M})$, $i=1, \cdots, M-1$, where $z_j\equiv\alpha_j$ and $z_{j+M}\equiv\alpha_j^*$ ($i=1, \ldots, M$) will be taken as $2(M-1)$ independent variables ($z_1=z_{M+1}\equiv 1$). Again, this will only imply that when we now set 
\begin{eqnarray}
F_i(z_1,  \ldots, z_{2M}) &=& 0,
\label{minors}
\end{eqnarray}
for $i=1, \cdots, M-1$, some of the solutions we find do not correspond to product vectors, although all the $\ket{e_i,f_i}$ we look for are among the solutions of (\ref{minors}).

 We have $2(M-1)$ variables $z_i$ and the same number $2(M-1)$ of polynomial equations for them, $M-1$ coming from the minors $F_i(z_1, z_{2M}) = 0$ and the remaining $M-1$ from its complex conjugation, which are inequivalent to the first ones as the variables are mapped according to $z_i \leftrightarrow z_{i+M}$, $i = 1,\cdots,~M$, under complex conjugation.

 No theorem exists for complex polynomials $P(\vec{\alpha}, \vec{\alpha}^*)$ which allows us to know the number of roots they have. However, in a {\em generic} case, namely when $P(\vec{\alpha}, \vec{\alpha}^*)$ is not proportional to its complex conjugate, we can prove that only a finite number of solutions exist. In \cite{MCK} a method to find such roots was developed for polynomials depending on one $\alpha$ and its complex conjugate. Accordingly, from $P(\alpha,\alpha^*)$ another polynomial $Q(\alpha)$ containing all the roots of $P$ was obtained. Such a method admits a generalization to the present case, which we shall discuss later on by means of an example when analyzing states of a $3\times 3$ system. As already mentioned, we were not able to determine when a density matrix $\varrho$ will lead to a set of {\em non-generic} polynomials. However, we expect this to be rarely the case. 
In what follows we will assume that the polynomials derived from $\varrho$ are {\em generic}, and that therefore there is only a finite number of product vectors that can appear in (\ref{separable}).

\subsection{Separability criterion.}

When the number of solutions of equation (\ref{conditions}) is finite, we 
can formulate a necessary and sufficient separability condition which follows from the following general theorem:  

{\bf Theorem 2} (see also \cite{tran}) {\bf .-}
A state $\varrho$ of rank $r(\varrho)$ is separable iff it can be written as a convex combination of at most $\min \{r(\varrho)^2,r(\varrho^{T_A})^2\}$ {\em linearly independent} projectors $\proj{e_i,f_i}$ onto product vectors.

{\it Proof.-} The inverse implication is obvious. For the direct implication we will assume, without loss of generality, that $ r(\varrho) \leq r(\varrho^{T_A})$. Caratheodoris' theorem tells us then that $\varrho$ can be expressed as a convex combination of $r(\varrho)^2$ product projectors,
\begin{equation}
\varrho = \sum_{i=1}^{r(\varrho)^2} p_i \proj{e_i,f_i}.
\end{equation}
Suppose these projectors are not linearly independent. This means we can find $\sum_i c_i \proj{e_i,f_i} = 0$ with at least some non-vanishing $c_i \in {\cal R}$. Set $\lambda \equiv \min \{p_i/c_i\}$. Then the decomposition 
\begin{equation}
\varrho = \sum_{i=1}^{r(\varrho)^2} (p_i-\lambda c_i) \proj{e_i, f_i},
\end{equation}
corresponds also to a convex combination of the previous projectors $\proj{e_i, f_i}$, but with at least one of the terms having vanishing weight. Now, if the remaining projectors do not form yet a linearly independent set, we can repeat the same procedure and get rid of another product projector. This can be iterated until expressing $\varrho$ as a convex combination of linearly independent product projectors. $\Box$

 Consequently, once we obtain all product vectors $|e_i,f_i\rangle\in R(\varrho)$ such that $|e_i^*,f_i\rangle \in R(\varrho^{T_A})$, $i=1, \cdots, L < \infty$, we can find out whether $\varrho$ is separable by proceeding as follows: 
\begin{itemize}

\item We build all possible maximal subsets of linearly independent projectors $\proj{e_i, f_i}$ (with at least $L_0\equiv\max\{r(\varrho), r(\varrho^{T_A})\}$ elements). Notice that there is only a finite number of subsets.

\item  For each of these subsets we express $\varrho$ as a linear combination of projectors in the subset. 

\item If this is possible, then we have to see whether the coefficients of the linear combination are all positive. 

\end{itemize}

We immediately have:

{\bf Separability criterion.-} $\varrho$ is separable iff all coefficients are non-negative in (at least) one of the linear combinations described above.

\subsection{Numerical methods.}

We notice that for a $\varrho$ with just a finite, but large number $L$ of eligible product vectors it may be impractical to construct all possible subsets of linearly independent product projectors, as described above. In this case the linear programming theory \cite{farkas} has developed various methods to try to find out a solution to whether $\varrho$ can be expressed as a linear combination, with positive weights, of the over complete but finite set of projectors $\proj{e_i, f_i}$.
We propose,  however, to use
for this aim the best separable approximation (BSA) method, 
developed by us in Ref. \cite{M&A}. It has nice physical analogies
also for non-separable states, providing the expansion
$$\varrho=\varrho_s+(1-\lambda)\delta\varrho,$$
where $\varrho_s=\sum_i\Lambda_iP_i$ is a separable state, 
$\lambda=\sum_i\Lambda_i$ 
is maximal, and finally $\delta\varrho$ is a state that does 
not have any product vector in its range. The paper \cite{M&A} 
describes an efficient algorithm for finding such expansion, by optimizing
each of the $\Lambda_i$ individually, and each of the pairs $\Lambda_i$, 
$\Lambda_j$ with respect to $\Lambda_i+\Lambda_j$. For the purpose 
of checking if a given matrix is separable, the BSA method of Ref. \cite{M&A} is sufficient; in the context of the present paper it is 
interesting to introduce here a generalization of the results of  \cite{M&A}
to the PPT states \cite{inprep}:

{\bf Lemma 7.-} Let $\varrho$ be a PPT state.
 For a given set of $P_i=|e_i,f_i\rangle\langle e_i,f_i|$, 
such that the  product vectors $|e_i,f_i\rangle 
\in R(\delta\varrho)$, such that $|e^*_i,f_i\rangle 
\in R(\delta\varrho^{T_A})$,  there exists the best 
separable approximation of $\varrho$, in the form 
 $$\varrho=\varrho_s+(1-\lambda)\delta\varrho,$$
where  $\varrho_s=\sum_i\Lambda_iP_i$ is a separable state, 
$\lambda=\sum_i\Lambda_i$ 
is maximal, and finally both $\delta\varrho\ge 0$, 
and $\delta\varrho^{T_A}\ge 0$. Moreover, there does not 
exist a product vector $|e,f\rangle 
\in R(\delta\varrho)$, such that $|e^*,f\rangle 
\in R(\delta\varrho^{T_A})$.

The proof of the above lemma is the same as the proof in Ref. \cite{M&A}. 
Similarly, one can find an efficient algorithm for finding the BSA, by requiring
 that:
\begin{itemize}

\item  All $\Lambda_i$ should be maximal, i.e.
\begin{eqnarray}
  \Lambda_i={\rm min}\left( \langle  e_i,f_i|(\rho-\sum_{j\ne i}\Lambda_j
P_j)^{-1}|e_i, f_i\rangle^{-1}, \right. \nonumber \\
\left.\langle  e^*_i,f_i|(\rho^{T_A}-\sum_{j\ne i}
\Lambda_jP^{T_A}_j)^{-1}|e^*_i, f_i\rangle^{-1}\right).
\end{eqnarray}

\item All pairs of $\Lambda_i,\Lambda_j$ should be maximized with 
respect to $\Lambda_i+\Lambda_j$. This requirement can also be expressed
in an analytical
form  for $\Lambda$'s, which will be presented elsewhere \cite{karnas}.
\end{itemize}

\subsection{Example: $3\times 3$ system.}

We end this section by describing with an example in a $3\times 3$ system how to estimate the number $L$
 of eligible product vectors. This example illustrates how to generalize to several independent $\alpha$'s the method developed in \cite{MCK}.

Suppose $r(\varrho)\leq 4$ and $r(\varrho^{T_A}) \leq 9$. For  $r(\varrho)= 4$ and $r(\varrho^{T_A})=9$ (least favorable case) we have that the matrix $A$ reads
\begin{eqnarray}
A =\left[ \begin{array}{c}
         \bra{k_1^1} + \alpha_2 \bra{k_1^2} +\alpha_3 \bra{k_1^3}  \\
         \bra{k_2^1} + \alpha_2 \bra{k_2^2} +\alpha_3 \bra{k_2^3}  \\
         \bra{k_3^1} + \alpha_2 \bra{k_3^2} +\alpha_3 \bra{k_3^3}  \\  
         \bra{k_4^1} + \alpha_2 \bra{k_4^2} +\alpha_3 \bra{k_4^3}  \\
         \bra{k_5^1} + \alpha_2 \bra{k_5^2} +\alpha_3 \bra{k_5^3} 
       \end{array}
      \right ], \nonumber \\ 
\end{eqnarray}
so that after constructing the $3\times 3$ submatrices $A_{1,2,3}$ by taking the first two rows of $A$ and one of the remaining rows at a time, we obtain three 3-rd order equations for $\alpha_1$ and $\alpha_2$:
\begin{eqnarray}
F_1 &=& \det M_1 \equiv \sum_{k=0}^3 \alpha_2^k P^k_{3}(\alpha_3) = 0, \label{f1} \\
F_2 &=& \det M_2  \equiv \sum_{k=0}^3 \alpha_2^k Q^k_{3}(\alpha_3)= 0, \label{f2} \\
F_3 &=& \det M_3  = 0,
\label{f33}
\end{eqnarray}
where $P_{s}(x)$ denotes a $s$-th order polynomial in $x$.  
By only using equations (\ref{f1}) and (\ref{f2}) we can obtain two quadratic equations in $\alpha_2$ as follows: on the one hand we multiply (\ref{f1}) by $Q^3_3(\alpha_3)$,  (\ref{f2}) by $P^3_3(\alpha_3)$, and then subtract them, leading to 
\begin{equation}
\sum_{k=0}^2 \alpha_2^k R_6^k(\alpha_3) = 0;
\label{f3}
\end{equation}
 on the other hand we multiply (\ref{f1}) by $Q^0_3(\alpha_3)$, (\ref{f2}) by $P^0_3(\alpha_3)$, again subtract them, and after dividing by $\alpha_2$ we obtain
\begin{equation}
\sum_{k=0}^2 \alpha_2^k S_6^k(\alpha_3) = 0.
\label{f4}
\end{equation}
Finally, applying the same trick but now to equations (\ref{f3}) and (\ref{f4}), we obtain two linear equations for $\alpha_2$, from which a unique $18$-th order equation for $\alpha_3$ follows. Therefore there are at most $18$ different values of $\alpha_3$ which in principle could lead to an eligible product vector. For each such values one should now still solve the $3$ $3$rd order equations (\ref{f1}-\ref{f33}) for $\alpha_2$, and see which solutions survive, if any 
\footnote{Notice that for a given $\alpha_3$ in principle we could find $0,1,2$ or $3$ valid values of $\alpha_2$. For simplicity we assume in the final estimation of the number $L$ of eligible product vectors that to each solution $\alpha_3$ there corresponds at most one valid $\alpha_2$.}
. Finally, for those triads $[1, \alpha_2, \alpha_3]$ which indeed fulfill (\ref{f1}-\ref{f33}) we can diagonalize $A$ and find the corresponding Bob's local vector $\ket{f}$ in the kernel of $A$. We have obtained, thus, $L \leq 18$.

\vspace{2mm}

Before going into the conclusions we shall discuss briefly the question of the relative size of $r(\varrho)$ and $r(\varrho^{T_{A}})$.
It is natural to expect that this difference is not too big.
However some naive intuitions must be abandoned (see \cite{barely}).
Here we shall make the simple observation:

{\bf Observation .-} Let $\varrho$ be a PPT state.
If kernel of $\varrho$ contains the range of
some PPT state $\sigma$, then the kernel of $\varrho^{T_{A}}$
contains the range of $\sigma^{T_{A}}$, so that $r(\varrho^{T_{A}})\leq MN-r(\sigma^{T_{A}})$.

The above observation about rank of $\varrho$
follows easily from the fact that Tr$(AB)=$Tr$(A^{T_{A}}B^{T_{A}})$.
Note that $\sigma$ can be a separable state.
In particular, if the kernel of $\varrho$
contains any system of $n$ orthogonal product
vectors (in particular UPB set \cite{UPB}) then
$r(\varrho^{T_{A}})$ can not exceed the value
of $MN-n$.
The same holds if $\sigma$ from our observation is
PPT bound entangled state defined as a UPB complement \cite{UPB}.
The rank of the latter does not change under partial transpose, so again
$r(\varrho^{T_{A}})$ can not exceed the value
of $MN-r(\sigma)$. It can be also extended in other direction:
taking $\sigma$ as a nontrivial PPT invariant state.
Apart from all $\sigma$'s being complements of real UPB's,
there is an other nontrivial class (provided in \cite{contin}) of
$N\times N$ states of that kind all having
$r(\sigma)=\frac{N(N-1)}{2}+1$. From the above discussion and the Theorem 1 we immediately
know, for example,  that for all the $3 \times 3$ PPT entangled states
with the kernel containing UPB complement
both ranks: $r(\varrho^{T_{2}})$ and $r(\varrho^{T_{2}})$
must amount to either $4$ or $5$ so
they cannot differ much from each other.

\section{Conclusions}
We have presented in this paper a relatively complete list 
of separability criteria for density matrices of low rank.
There are several problems, however, which remain open and 
are worth further studies:

\begin{itemize} 

\item In our analysis of the kernels of $\varrho$
 and $\varrho^{T_A}$
we have  essentially used only those of their properties that are consequences
of the dimensionality. On the other hand, it is expectable that both
kernels are structurally related through the partial transpose operation.
It would be important to investigate such relations, since it would probably 
automatically put much more stringent restrictions on the existence 
of separable matrices of low rank. 

\item All of the results of this paper can be generalized to the 
case of multipartite systems, and in particular $3$ 
partite systems. We have already obtained several results,
 but we leave a detailed and complete discussion of this 
problem to a separate publication. Let us just mention here that
according to our studies:
i) there are no rank $N$ 
PPT entangled states for $N\times N\times N$
systems; ii)  In $2\times 2\times 2$ spaces  PPT states of rank 4
are separable with respect to $2\times 4$ space of
 Alice and joint space of Bob and Charles,       
and posses a unique decomposition into a sum of 4
projectors onto product 
vectors in $2\times 4$ space; they are fully separable iff 
those 4 product vectors are  at the same time product vectors in the 
sense of $2\times 2\times 2$; iii)  In $2\times 2\times 2$ spaces  generic 
PPT states
with $r(\varrho)+r(\varrho^{T_A})+r(\varrho^{T_B})+r(\varrho^{T_C})\le
4\times 8-2\times 2 + 1=29$ have a finite number of product vectors in
their range, such that the partial conjugates of those product 
vectors are in the corresponding ranges of partial transposes.

\end{itemize}

This work has been supported by Deutsche Forshungsgemeinschaft
(SFB 407 and  Schwerpunkt ``Quanteninformationsverarbeitung''), the \"Osterreichisher Fonds zur F\"rderung der
wissenschaftlichen Forschung (SFB P11), the European TMR network
ERB-FMRX-CT96-0087, and the Institute for Quantum Information
Gmbh. J. I. C. thanks the University of Hannover for
hospitality. P. H. acknowledges the
grant from Deutscher Akademisher Austauschdienst.
We thank S. Karnas, A. Sanpera, J. Smolin, B. Terhal for fruitful
discussions. We thank J. Werner for indicating to us 
relations to linear programming theory.


\begin{references}
\bibitem{truco} Let $\ket{\Psi_f} \equiv \left[\begin{array}{c} \ket{f}\\ -F\ket{f} \end{array}\right]$ and $M\equiv \left[\begin{array}{cc}  E & F^{\dagger}\\F & I \end{array}\right] \geq 0$, where $\ket{f} \in {\cal C}^N$ and $E, F, I$ are operators on ${\cal C}^N$.  Then $\forall \ket{f}$ we have $\bra{\Psi_f} M\ket{\Psi_f} = \bra{f} E - FF^{\dagger}\ket{f} \geq 0$.
\end{references}
\end{document}